# Microwave Photonics for Space-Ground Connectivity

Ruiqi Zheng[1,2], Jingxu Chen[1,2], Jinkun Hu[1,2], Haikun Huang[1,2], Junyi Zhang[1,2], Wufei Zhou[1,2], Sheng Dong[1,2], Xudong Wang[1,2], Xinhuan Feng[1,2], Jiejun Zhang[1,2*], Jianping Yao[1,3*]

1. International Cooperation Joint Laboratory for Optoelectronic Hybrid Integrated Circuits, Jinan University, Guangzhou, 511443, China,
2. College of Physics & Optoelectronic Engineering, Jinan University, Guangzhou 510632, China
3. Microwave Photonics Research Laboratory, School of Electrical Engineering and Computer Science, University of Ottawa, Ottawa, ON K1N 6N5, Canada
*Correspondence to: zhangjiejun@jnu.edu.cn, jpyao@uottawa.ca

**Abstract:** Future space-ground communication networks require a seamless fusion of technologies that combine the all-weather reliability of microwave links with the ultra-high data capacity of near-infrared optical systems. Achieving this vision demands compact, robust, and multifunctional hardware, yet monolithic integration of these fundamentally distinct domains has remained elusive. Here, we present the first monolithically integrated silicon photonic chip that bridges microwave and optical domains for dual-band free-space communications and dynamic beamforming. The chip integrates a microwave true time delay (TTD) beamforming network, an optical phased array (OPA) beamforming network, and an optical coherent transceiver, all on a silicon-on-insulator (SOI) platform. By uniting the strengths of microwave resilience, optical bandwidth, and coherent detection sensitivity, this photonic integrated circuit represents a critical step toward reconfigurable, interference-resistant, high-throughput links for satellites, UAVs, and ground stations. Experimental demonstrations confirm two-dimensional dynamic beam steering in both bands 24.9°×18.5° at microwave frequencies and 10°×4.7° in the optical domain. In a 5-meter free-space link, the chip achieves error-free transmission at 10 Gb/s for microwave and 80 Gb/s per wavelength in the near-infrared band. These results establish integrated microwave photonics as a promising platform for bridging Earth and orbit through compact, dual-band, beamforming-enabled transceivers.

## 1. Introduction

In recent years, low Earth orbit (LEO) satellite constellations, exemplified by SpaceX's Starlink, have sparked a global revolution in space communication technologies [1-3]. By establishing high-speed communication infrastructure with global coverage, LEO satellite networks are poised to become a key enabler of the integrated space-terrestrial network envisioned for 6G [1-3]. Compared to traditional geostationary satellites, LEO satellites offer distinct advantages including low transmission latency and reduced path loss. However, their high-speed mobility introduces critical challenges such as dynamic topology network management and the need for rapid, precise beam alignment [4-6]. To address these challenges, robust beamforming solutions capable of real-time adaptation to maintain reliable communication links must be found. Therefore, multi-band cooperative communication, which synergizes the high reliability of microwave links with the large bandwidth of optical links, has emerged as a pivotal solution to overcome bottlenecks in space-air-ground communication networks [7, 8].

Microwave phased arrays [9-16] and optical phased arrays (OPA) [17-26] are the foundational beamforming technologies for space-air-ground communications, each offering distinct advantages: microwave systems provide robust, all-weather connectivity, while optical systems offer ultra-large bandwidth for high-capacity data links. The integration of these complementary technologies presents a transformative approach to address the evolving demands of next-generation space-air-ground communication networks. The microwave band remains the workhorse for current inter-satellite and satellite-ground communications [27-29], while emerging near-infrared photonic systems enable revolutionary terabit-per-second-capacity laser inter-satellite connections [30-32]. This dual-band approach combines the reliability of established microwave technology with the transformative capacity of optical communications. Existing studies have demonstrated that microwave-optical hybrid systems can enhance robustness while maintaining a high data rate space networks [7, 8]. However, conventional space-air-ground communication networks are implemented based on discrete components, which suffer from large size, high weight, and excessive power consumption, making them unsuitable for satellite payloads with stringent integration and energy efficiency requirements [24, 27].

Silicon photonic (SiP) integration is considered an effective solution for implementing miniaturized dual-band beamforming and simultaneous free-space microwave and optical communications. In addition to its CMOS-compatibility, the high refractive index contrast and large-scale integration capability make SiP particularly well-suited for the implementation of space-air-ground multi-band transceiver systems [23-25, 33-36], enabling the integration of high-speed electro-optic modulators, low-loss waveguide networks, and high-performance photodetectors (PDs) [27, 29-32, 37-40]. Although prior studies have demonstrated independent functionalities such as true time delay (TTD) microwave beamforming [9-16], OPA beamforming [17-26], and optical coherent detection [30, 41-46], no work has yet reported to realize a monolithically integrated dual-band microwave and near-infrared transceiver for beamforming and simultaneous free-space microwave and optical communications.

This study proposes and experimentally demonstrates the world's first monolithically integrated silicon photonic chip enabling dual-band microwave and near-infrared beamforming and simultaneous free-space microwave and optical communications.

The chip integrates a microwave TTD beamforming network, an OPA system, and an optical coherent transceiver to combine the high connection reliability of microwave communication with the large bandwidth of near-infrared communication, and with the high-sensitivity detection capabilities and environmental interference resistance of the optical coherent communication. The microwave TTD beamforming network is implemented using four-channel 3-bit optical switch delay lines (OSDLs) with integrated PDs. The OPA is implemented using sixteen-channel thermo-optic delay lines (TODLs) with integrated waveguide grating antennas. The optical coherent transceiver is implemented using a dual-parallel Mach-Zehnder modulator (DPMZM), a 90° optical hybrid and four integrated PDs. The chip is fabricated and experimentally tested. The experimental results show that the chip achieves two-dimensional dynamic beam steering control, with microwave beam scanning over 24.9°×18.5° and optical beam scanning over 10°×4.7°. In a 5 m free-space communication link, microwave communication of 5 GBaud-QPSK signal modulated on a microwave carrier at 30 GHz with an error vector magnitude (EVM) of 21.9% is achieved, while near-infrared communication of 20 GBaud-16QAM with an EVM of 11.3% is realized, fully validating the good performance of the chip for dual-band communications. By synergistically optimizing the microwave and near-infrared functionalities, the chip ensures reliable connectivity while enhancing transmission bandwidth, paving the way for lightweight, energy-efficient LEO satellite networks.

## 2. Dual-Band Silicon Photonic Transceiver Design

A space-terrestrial communication system supporting dual-band free-space optical and microwave links is illustrated in Fig. 1(a). Ultra-high-capacity optical links are used for intersatellite communication, while all-weather-reliable microwave links support communications between satellites and ground stations. To enable both optical and microwave beamforming as well as dual-band communications, a monolithically integrated silicon photonic chip is developed. This chip innovatively combines three subsystems on a single chip: a microwave TTD beamforming network, an OPA system, and an optical coherent transceiver. The architecture of the chip and its micrograph are shown in Fig. 1(b) and (c).

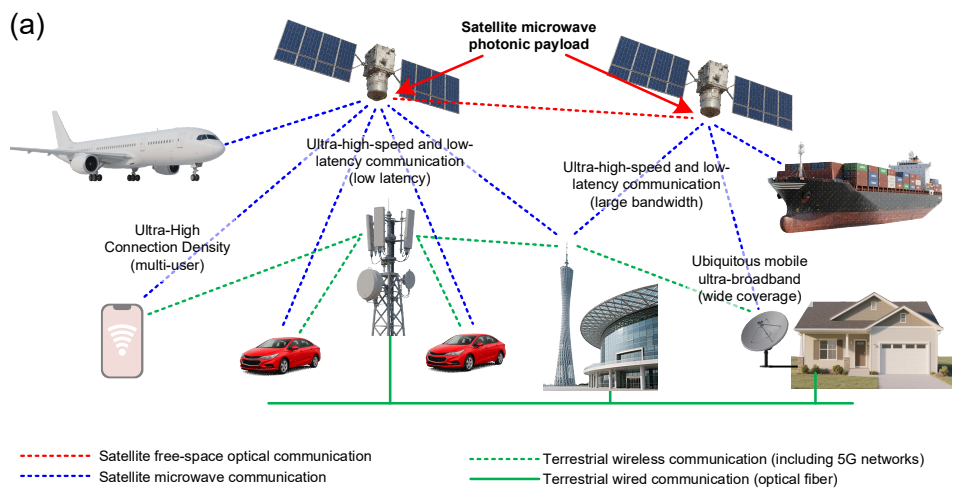

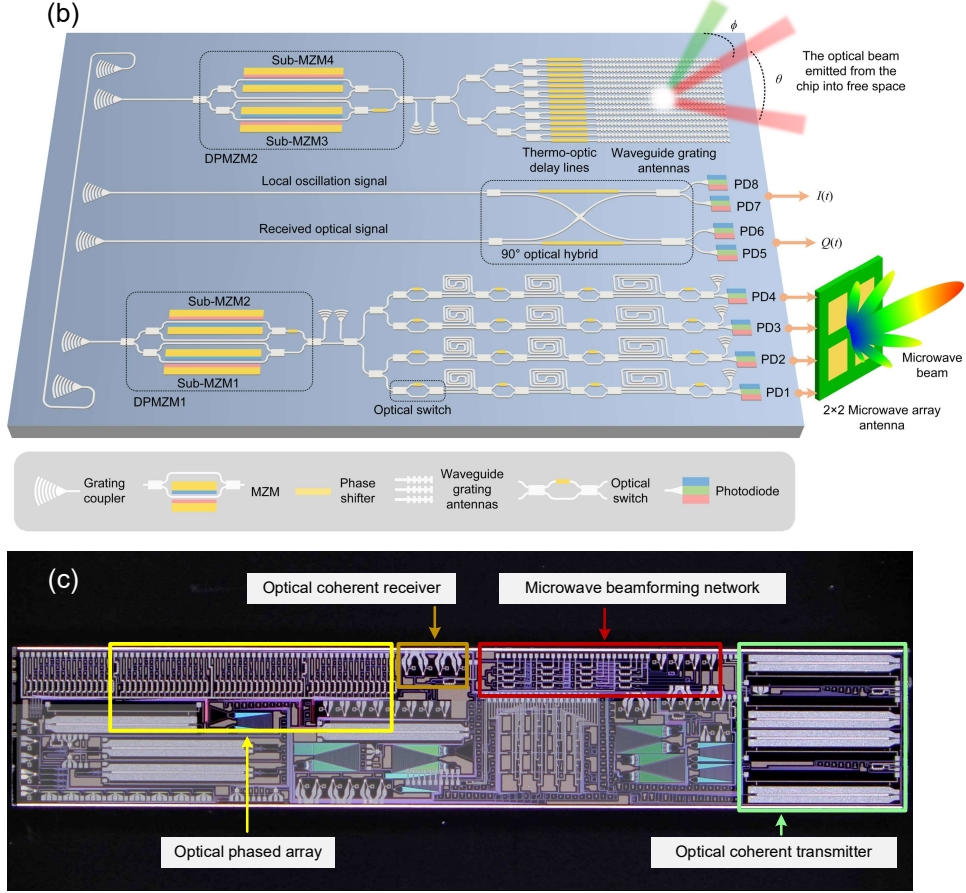

**Fig. 1.** (a) A space-terrestrial communication system supporting dual-band free-space microwave and optical links. (b) The schematic diagram of the transceiver incorporating a microwave TTD beamforming network, an OPA network, and an optical coherent transceiver. (c) The micrograph of the silicon photonic chip.

## 2.1. Microwave TTD Beamforming Network

As shown in Fig. 1, the microwave TTD beamforming network consists of an off-chip laser source, an on-chip modulator, four on-chip 3-bit OSDLs, and four on-chip PDs. First, the off-chip laser source emits a continuous wave (CW) light serving as an optical carrier that is coupled into the chip via a grating coupler. The light is then fed into the on-chip modulator where a microwave signal is modulated on the optical carrier. Subsequently, a two-stage 1×2 multimode interferometer (MMI) serving as a 1×4 optical coupler is used to split the optical signal into four paths. These four optical signals are routed through the four parallel 3-bit OSDLs. Each OSDL comprises four Mach-Zehnder interferometer (MZI)-based optical switches and three waveguide delay segments with delays of $\Delta\tau$, $2\Delta\tau$, and $4\Delta\tau$. By controlling the switching states of these four optical switches, the input optical signal can be delayed from 0 to $7\Delta\tau$ with a step of $\Delta\tau$. The fourth-stage optical switch in each delay line has two output ports, with one being used to deliver the delayed optical signal to the PD for optoelectronic conversion and the other as a calibration port for measuring the switching voltages of all four optical switches in the delay line. Additionally, the fourth-stage optical switch balances the optical power across all the four paths, ensuring uniform microwave signal power output for optimized beamforming performance. Finally, the four delayed optical signals are converted to microwave signals by the four integrated PDs. These microwave signals are then transmitted via RF cables to four off-chip microwave amplifiers and 2×2 microwave antenna array. After amplification, the microwave antenna array radiates the time-delayed microwave signals into free space. To form a microwave beam at a desired steering angle, the microwave signals emitted by adjacent-row or adjacent-column antennas must satisfy

$$\theta = \sin^{-1}\left(\frac{c\Delta\tau}{d}\right) \quad (1)$$

where $\theta$ is the horizontal or vertical steering angle of the microwave beam, $c$ is the speed of light in vacuum, $\Delta\tau$ is the time delay difference between microwave signals emitted by adjacent-row or adjacent-column antennas, $d$ is the antenna array spacing. As given in Eq. (1), the beam steering direction in the two-dimensional space can be controlled by tuning the delays of the four on-chip OSDLs.

**2.2. Optical Phased Array Network**

As shown in Fig. 1, the OPA beamforming network consists of an off-chip laser source, a 16-channel on-chip TODLs, and a 16-channel on-chip waveguide grating antenna array. First, the off-chip laser source emits a CW light, which is coupled into the chip via a grating coupler. The light is then split into 16 channels using a cascaded four-stage 1×2 MMI splitter network and subsequently routed to the 16 TODLs. These 16 optical signals after experiencing different delays are then sent to the waveguide grating antenna array and radiated into free space. For a grating antenna element with a grating period $\Lambda$, the phase-matching condition for an incident optical signal diffracted into free space at an angle $\theta_{air}$ is given by

$$\frac{2\pi n_{eff}}{\lambda}\Lambda - \frac{2\pi n_{air}}{\lambda}\Lambda \sin(\theta_{air}) = 2\pi N \quad (2)$$

where $n_{eff}$ is the effective refractive index of the waveguide grating, $n_{air} = 1$ is the refractive index of air, $\theta_{air}$ is the diffraction angle, and $N$ is the diffraction order. Equation (2) shows that the diffraction angle $\theta_{air}$ depends on $\lambda$, enabling the elevation angle of the beam steering through laser wavelength tuning. The azimuth angle of the beam is steered by adjusting the voltages applied to the TODLs. Adjusting the heating power modifies the effective refractive index of the optical waveguide, thereby tuning the optical time delays. Similar to the microwave TTD beamforming, when the time delay between adjacent channels satisfies Eq. (1), the 16 optical signals radiated by the waveguide grating antennas coherently form a laser beam at the desired azimuth angle. Therefore, by jointly controlling the laser wavelength and the relative time delays among the 16 optical channels, the two-dimensional beam steering of the laser beam can be achieved.

## 2.3. Optical Coherent Transceiver

As shown in Fig. 1, the optical coherent transceiver integrates an on-chip modulator connected to the OPA system, two balanced photodetectors (BPD), and a 90° optical hybrid. The on-chip modulator employs a DPMZM configuration, comprising two Mach-Zehnder modulators (MZMs) and a thermal phase shifter. Baseband in-phase (I) and quadrature (Q) communication signals are modulated on the optical carrier at the respective MZMs within the DPMZM, while the thermal phase shifter enforces a 90° phase shift between the two optical paths to achieve quadrature superposition. The 90° optical hybrid consists of two 1×2 MMIs, two 2×2 MMIs, two thermal phase shifters, and a waveguide crossing. One input port receives the free-space optical communication signal, while the other accepts the local oscillator (LO) light for coherent detection. The two thermal phase shifters independently adjust the phase difference between the received signal and LO to satisfy the phase condition of coherent demodulation requirements. Through precise phase control, the hybrid generates four optical signals at the four output ports given by: $E_s+E_{LO}$, $E_s-E_{LO}$, $E_s+jE_{LO}$, and $E_s-jE_{LO}$, where $E_s$ represents the received optical signal and $E_{LO}$ denotes the LO signal. These four optical signals are fed into two BPDs, each containing two PDs. The balanced output is obtained by subtracting the photocurrents from the PD pair, either through off-chip circuitry or integrated differential amplifiers. The first BPD processes $E_s+E_{LO}$ and $E_s-E_{LO}$ to recover the in-phase (I) baseband signal, and the second BPD processes $E_s+jE_{LO}$ and $E_s-jE_{LO}$ to extract the quadrature (Q) baseband signal. By dynamically tuning the thermal phase shifters in the 90° hybrid, the system ensures orthogonal demodulation, enabling simultaneous recovery of the I and Q components for coherent communication.

## 3. Experimental Results

To experimentally validate the operation of the proposed dual-band transceiver chip and evaluate its performance, we first implement the optical and electrical packaging of the chip and integrate it into a compact chassis, as shown in Fig. 2. Detailed packaging specifications are provided in Supplementary Material S2.

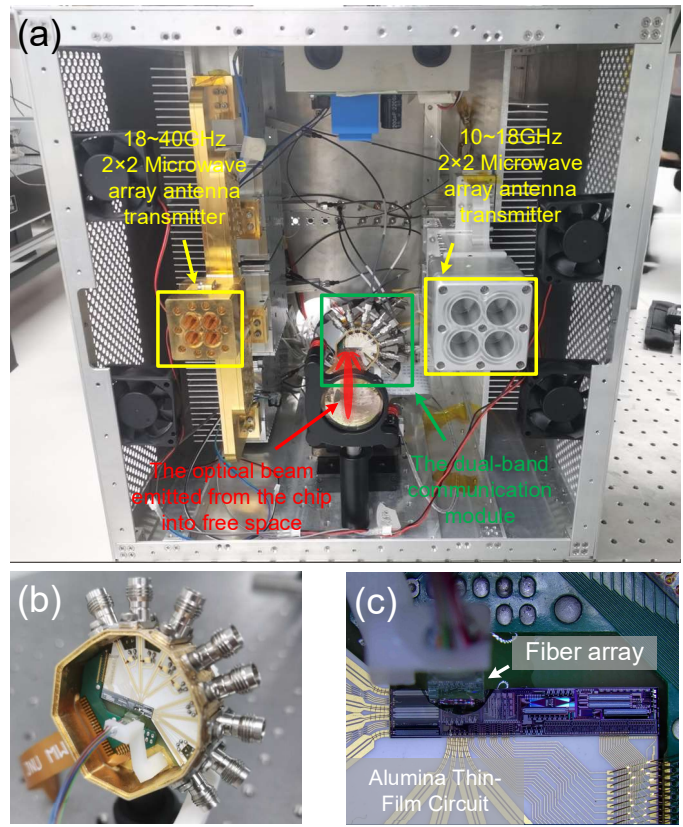

**Fig. 2.** (a) Packaged dual-band silicon photonic transmitter. The green box highlights the module after optical and electrical packaging, and the yellow boxes highlight the microwave array antennae for beamforming and free-space transmission. (b) The dual-band silicon photonic transmitter module after optical and electrical packaging. (c) Microscopic image of the internal structure of the silicon photonic transmitter module.

**3.1. Two-Dimensional Microwave Beam Steering**

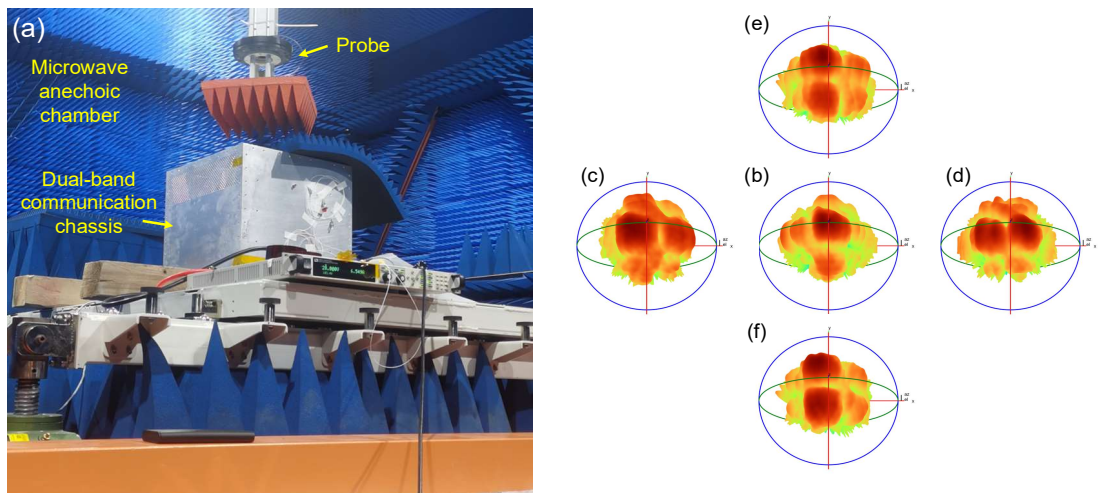

**Fig. 3.** (a) Measurement setup for microwave beam characterization in a microwave anechoic chamber. The measured microwave beams under different steering conditions: (b) beam with 0° steering angle, (c) beam steered leftward in the horizontal direction, (d) beam steered rightward in the horizontal direction, (e) beam steered upward in the vertical direction, (f) beam steered downward in the vertical direction.

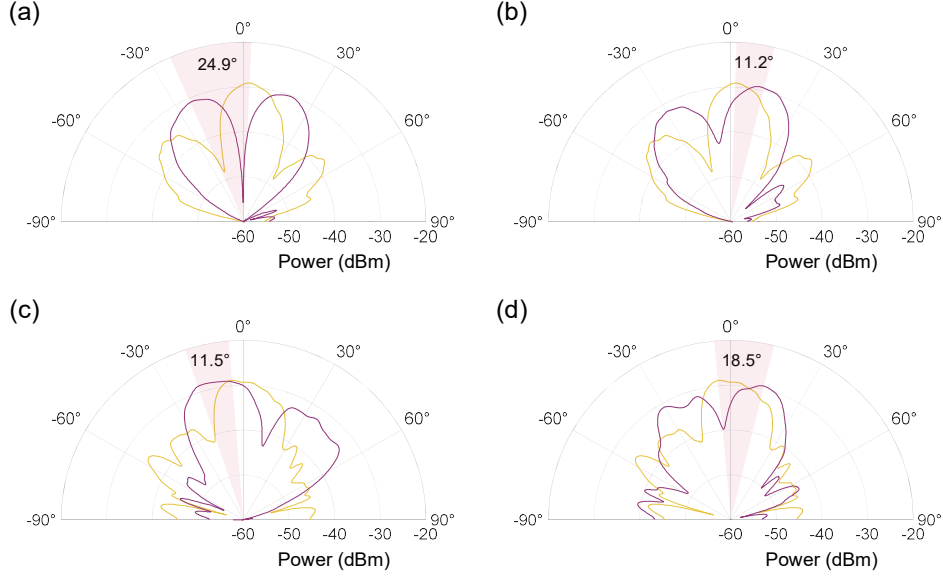

**Fig. 4.** Comparison of the beam energy distributions along the vertical cross section (a) between Fig. 3(b) and Fig. 3(f), (b) between Fig. 3(b) and Fig. 3(e). Comparison of the beam energy along the horizontal cross section (c) between Fig. 3(b) and Fig. 3(c), (d) between Fig. 3(b) and Fig. 3(d).

After completing the calibration of the microwave TTD beamforming network, a tunable time delay from 0 to 98 ps with a tuning step of 14 ps supporting a microwave signal with its frequency spanning from 10 MHz to 43.5 GHz is obtained as detailed in Supplementary Materials S3 and S4. Then, we connect the four on-chip PDs of the microwave TTD beamforming network to the RF ports of an external 2×2 microwave antenna array to validate the beam-steering capability of the chip in a microwave anechoic chamber, as shown in Fig. 3(a). The frequency of the microwave signal modulated on the optical carrier is 13 GHz or a wavelength of 23 mm. The array spacing of the 2×2 array antenna is 26 mm. Fig. 3(b) shows the far-field radiation pattern when the time delays of the four OSDLs are set to [0 ps 0 ps 0 ps 0 ps]. Figs. 3(c), (d), (e), and (f) show the measured microwave beam patterns after steering the beam in four directions: leftward, rightward, upward, and downward, respectively, when the time delays are set to [0 ps 42 ps 0 ps 42 ps], [42 ps 0 ps 42 ps 0 ps], [0 ps 0 ps 42 ps 42 ps], and [42 ps 42 ps 0 ps 0 ps], respectively. A comparison of the beam energy distributions along the vertical and horizontal cross section between Fig. 3(b) and Figs. 3(c), (d), (e), and (f) is shown in Fig. 4. According to theoretical calculations based on Eq. (1), the theoretical beam steering direction is 28° when the delay difference is 42 ps. The average error is 11.4°, which is attributed to unequal RF cable lengths between the PDs and antenna array. The experimental results in Fig. 3 and Fig. 4 verify the capability of the chip for 2D microwave beam scanning.

## 3.2. Two-Dimensional Laser Beam Steering

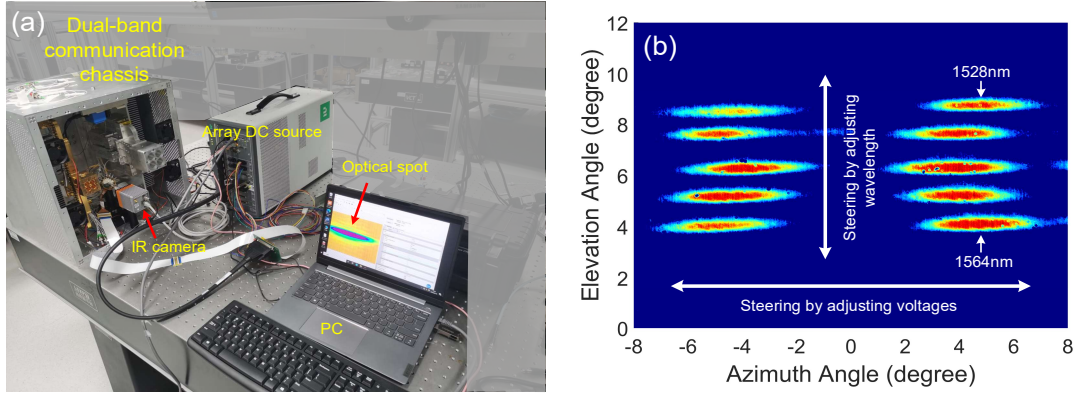

**Fig. 5**. (a) Measurement setup for optical beam characterization. (b) The measured two-dimensional laser beam scanning patterns.

After completing the calibration of the OPA system to focus the energy of the optical beam into a single spot as detailed in Supplementary Materials S5, the two-dimensional laser beam scanning pattern is measured, which is done by using an infrared (IR) camera to monitor the radiation pattern. First, wavelength tuning is used to steer the elevation angle of the beam, with the scanning angle given by Eq. (2). Due to the vertical fiber-coupling configuration and the fixed fiber array mount, the IR camera is positioned 70 mm away from the chip, as shown in Fig. 5(a). Figure 5(b) presents the beam scanning patterns under wavelength tuning. As the optical wavelength is tuned from 1528 nm to 1564 nm, the elevation angle decreases from 8.8° to 4.1°, yielding a tuning rate of 0.13° per nm. A broader wavelength tuning range would further extend the vertical beam scanning range. For azimuth angle steering, the voltages applied to the TODLs are adjusted to introduce optical delays between adjacent waveguide grating antennas according to Eq. (1). It can be seen in Fig. 5(b) that the azimuth angle of the laser beam can be steered from -5° to +5°. It should be noted that the measured azimuth angle scanning range is limited by the size of the charge-coupled device (CCD) in the IR camera. By repositioning the IR camera, the maximum horizontal scan range is measured to be ±60°. These measured beam patterns shown in Fig. 5 validate the beam steering capability of the OPA.

## 3.3. Dual-Band Communication Experiment

To validate its dual-band communication capability, an experiment based on the experimental setup shown in Fig. 6(a) is performed. Figure 6(b) is a photograph of the implemented hardware configuration. The distance between the transmitter and receiver is 5 m.

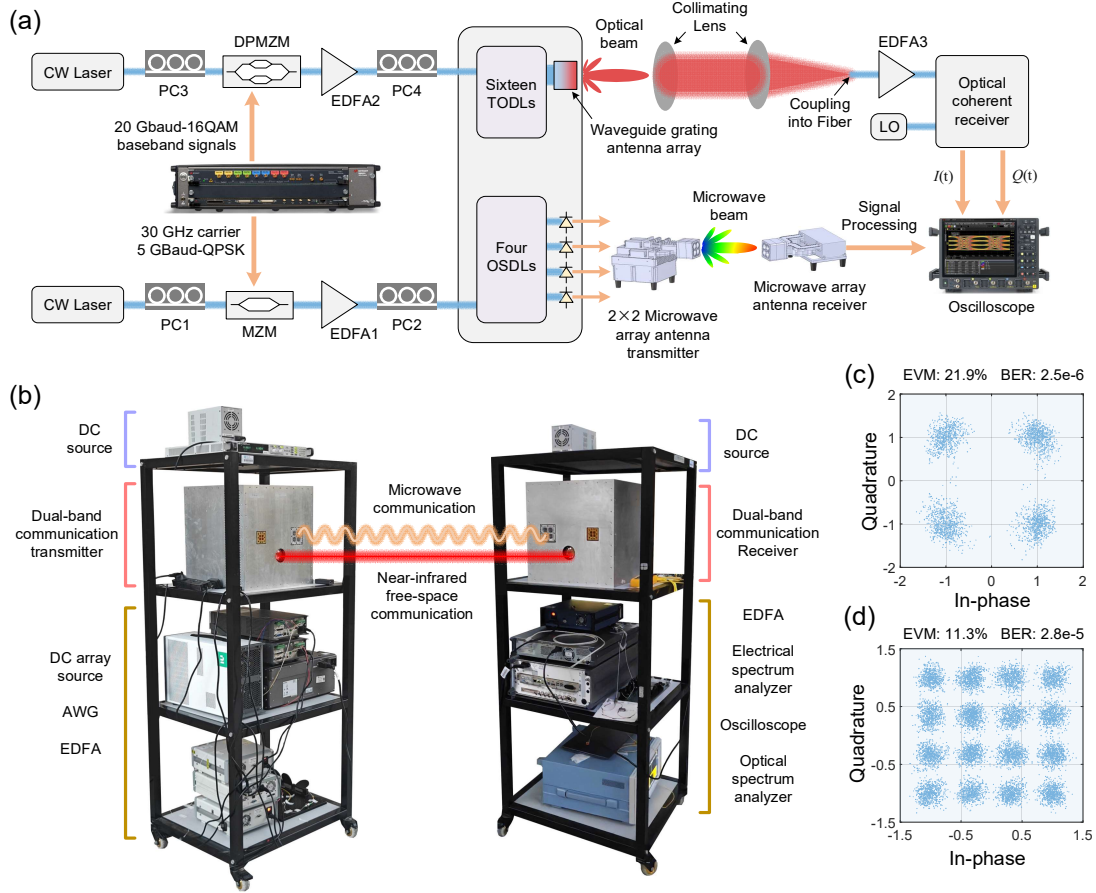

**Fig. 6**. (a) Schematic diagram of the experimental setup for dual-band microwave and optical communication. (b) Photograph of the implemented hardware configuration. (c) Constellation diagram of the received 5 GBaud QPSK signal transmitted via the microwave communication link. (d) Constellation diagram of the received 16-QAM signal transmitted via the IR communication link.

For microwave communication, the wavelength of the laser source (IDPHOTONICS CoBrite DX1) is tuned at 1560 nm with an output power of 15 dBm. This wavelength is within the calibrated wavelength range of the OSDLs. The optical carrier is sent to a Mach-Zehnder modulator (MZM) through a polarization controller (PC1), to which a 30 GHz microwave signal carrying 5 GBaud QPSK signal generated by an arbitrary waveform generator (AWG, Keysight-8194) is applied. The modulated optical signal is then amplified by an erbium-doped fiber amplifier (EDFA1) to compensate for the link loss and ensure sufficient link budget. A second PC (PC2) is used to adjust the polarization state to maximize the coupling of the light into the chip via a grating coupler. The modulated signal is processed by the on-chip microwave beamforming TTD network for delay control and radiated into free space via the 2×2 microwave antenna array. By programming the TTD beamforming network, the microwave beam is steered to precisely align with the receiver antenna. The received signal is captured by an oscilloscope (Keysight UXR0704A) for analysis. The constellation of the demodulated signal is shown in Fig. 6(c). The EVM is measured to be 21.9% and the Bit error rate (BER) is $2.5\times10^{-6}$. Error-free transmission can be achieved when forward error correction (FEC) is applied.

For near-infrared communication, the wavelength of the laser source is tuned at 1560 nm with an output power of 15 dBm. The beam is sent to a DPMZM through a PC (PC3). The DPMZM consists of two sub-MZMs, to which an in-phase (I) and a quadrature (Q) baseband signal with a π/2 phase shift between them are independently applied, enabling quadrature superposition. The I and Q baseband signals are generated by an AWG producing a 20 GBaud 16-QAM baseband waveform. Both the sub-MZMs are biased at the null point to maximize sideband power. The modulated optical signal is then amplified by an erbium-doped fiber amplifier (EDFA2) to compensate for the link loss and ensure an adequate optical power level for subsequent processing. The optical signal is then sent to the chip via a grating coupler through a second PC (PC4). By controlling PC4 to control the polarization of the incident light, the coupling efficiency to the chip is maximized. The signal is then processed by the on-chip optical phased array and emitted into free space via the optical array antenna. To prevent the beam from divergence, two collimating lenses are inserted to convert the beam into a parallel beam for long-distance transmission. At the receiver, the collimated beam is refocused into a single-mode fiber. To compensate for the free-space propagation loss and coupling losses, the received optical signal is amplified by an amplifier (EDFA3) before detection. The amplified signal is fed into the integrated coherent receiver. A LO laser source generating a continuous-wave light that has the same wavelength as that of the transmitter laser source is fed into the coherent receiver for coherent detection. The demodulated signal is analyzed using the oscilloscope. Figure 6(d) shows the constellation of the demodulated 16-QAM signal. The measured EVM is 11.3% and the BER is $2.8\times10^{-5}$. Again, error free transmission is enabled when forward error correction (FEC) is applied.

## 4. Conclusion

We have presented the world's first monolithically integrated silicon photonic chip capable of microwave and near-infrared dual-band beamforming and free-space microwave and optical communications. The chip consists of three key modules: a microwave TTD beamforming network, an OPA beamforming network, and an optical coherent transceiver. For microwave beamforming, four OSDLs were employed to achieve tunable time delays from 0 to 98 ps with a tuning step of 14 ps supporting a microwave signal with its frequency spanning from 10 MHz to 43.5 GHz. By independently controlling the four OSDLs, 2D beam steering of 24.9°×18.5° was achieved. For optical beamforming, 16 TODLs were employed to achieve tunable delays, and 16 waveguide gratings were connected to the TODLs as an optical antenna array. Through thermal tuning of the time delays combined with wavelength tuning, 2D laser beam steering of 10°×4.7° was achieved. In a 5-m free-space communication link, the microwave channel achieved 5 GBaud QPSK transmission at 30 GHz with 21.9% EVM, while the near-infrared channel supported 20 GBaud 16-QAM transmission with 11.3% EVM, both meeting error-free requirements when forward error correction (FEC) is applied. This work provides a highly integrated, high-performance silicon photonic solution for addressing multi-band cooperative communication challenges in LEO networks, demonstrating significant potential for enabling lightweight and energy-efficient communication architectures in future LEO satellite constellations.

## 5. Author contributions

R.Z. and J.C. contributed equally to this work. R.Z., J.Z. (Jiejun Zhang), and J.P.Y. conceived the idea. R.Z. and J.Z. (Jiejun Zhang) designed the chip layout. R.Z. and J.C. packaged the chip. R.Z., J.C., J.H., and H.H. conducted the experiments and analyzed the data. R.Z., J.Z. (Jiejun Zhang), and J.P.Y. wrote the manuscript. J.Z. (Junyi Zhang), W.Z., S.D., X.W., and X.F. contributed to the experiments.

## 6. Acknowledgements

This work was supported in part by the National Key Research and Development Program of China under Grant 2024YFB2807400, and in part by the Guangdong Engineering Technology Research Center for Integrated Space-Terrestrial Wireless Optical Communication.

# Supplementary Materials

# Microwave Photonics for Space-Ground Connectivity

Ruiqi Zheng[1,2], Jingxu Chen[1,2], Jinkun Hu[1,2], Haikun Huang[1,2], Junyi Zhang[1,2], Wufei Zhou[1,2], Sheng Dong[1,2], Xudong Wang[1,2], Xinhuan Feng[1,2], Jiejun Zhang[1,2*], Jianping Yao[1,3*]

1. International Cooperation Joint Laboratory for Optoelectronic Hybrid Integrated Circuits, Jinan University, Guangzhou, 511443, China,
2. College of Physics & Optoelectronic Engineering, Jinan University, Guangzhou 510632, China
3. Microwave Photonics Research Laboratory, School of Electrical Engineering and Computer Science, University of Ottawa, Ottawa, ON K1N 6N5, Canada
*Correspondence to: zhangjiejun@jnu.edu.cn, jpyao@uottawa.ca

# Contents

# S1. Fabrication Method

The silicon photonic chip was fabricated on a silicon-on-insulator (SOI) platform at the Advanced Micro Foundry (AMF). Key components including 14 grating couplers, 1×2 and 2×2 multimode interference (MMI) couplers, 36 thermo-optic phase shifters, 2 dual-parallel Mach-Zehnder modulators (DPMZM1 and DPMZM2) and 8 photodiodes (PDs) were designed using AMF's process design kit (PDK). The chip layout was designed using Luceda IPKISS photonic integrated circuit design software.

# S2. Optical and Electrical Packaging

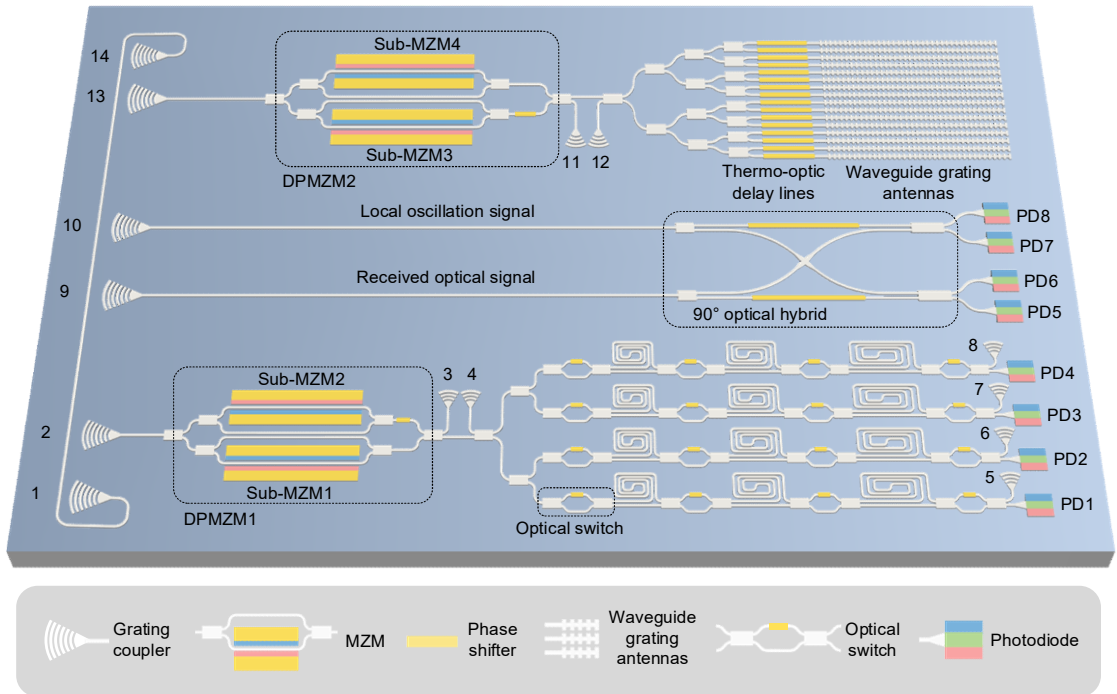

**Figure S1**. The schematic diagram of the transceiver incorporating a microwave TTD beamforming network, an OPA network and an optical coherent receiver for microwave and near-infrared dual-band free-space communication.

As shown in Fig. S1, the system employs 14 grating couplers for light coupling into or out of the chip (Ports 1 to 14). Port 1 and port 14, connected via a silicon waveguide, are used for measuring the coupling loss. Port 2 is used as the optical input port for the microwave true time delay (TTD) beamforming network. Port 3 is used as a test port for testing DPMZM1 (sub-MZM1 and sub-MZM2) within the microwave TTD beamforming network. Port 4 is used as an input port for calibrating the 4 optical switch delay lines (OSDLs), whereas ports 5~8 are used to perform delay lines calibration. Port 9 and port 10 are used as the input ports for the optical coherent receiver, receiving the communication optical signal and local oscillator (LO) signal, respectively. Port 13 is used as the optical input port for the optical phased array (OPA). Port 11 is used as a test port for testing DPMZM2 (MZM3 and MZM4) in the optical coherent transmitter. Port 12 is used as an input port for testing the OPA.

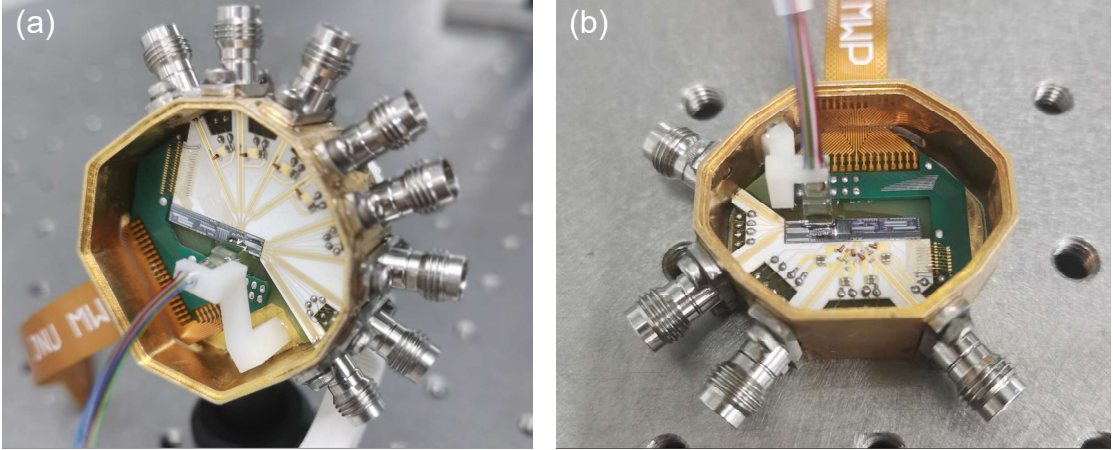

**Figure S2**. (a) Packaged dual-band transmitter including the microwave and optical beamforming networks. (b) Packaged coherent receiver.

The high-density integration of four modulators and eight PDs within the chip poses substantial challenges when consolidating the three subsystems within a single packaged module. This dense packaging leads to significant RF losses for both the modulators and the PDs, while also reducing their operational bandwidths. To ensure optimal performance of all on-chip components, we package the three subsystems into two separate modules, the transmitter module and the coherent receiver module. The transmitter module consists of the microwave TTD beamforming network and the OPA, as shown in Fig. S2(a). The receiver module has only the optical coherent receiver, as shown in Fig. S2(b). The optical packaging process primarily involves two key steps: 1) optimizing the alignment for high efficiency optical coupling and 2) UV-curing adhesive fixation. After packaging, we measure the coupling loss between port 1 and port 14, which are connected by a 2 mm waveguide. The total loss was 11 dB, indicating a coupling loss of approximately 5.5 dB per fiber-chip interface. Due to the high-density integration of multiple modulators and PDs on the chip, a high-dielectric-constant substrate is required to connect the RF electrodes on the chip to the external connectors while maintaining 50-Ω impedance matching. In the packaged modules, we employ alumina ($Al_2O_3$) as the substrate material, which offers a high dielectric constant $\varepsilon_r$ of 9.9 and an ultra-low loss tangent tan$\delta$ of 0.0001 for microwave signal transmission. The microstrip lines, fabricated using 2 μm gold traces, are optimized using HFSS simulations to ensure impedance matching and minimize signal reflection. To connect the RF electrodes to the microstrip lines, 25 μm diameter gold wire bonding is employed. Wire bonding is also used to connect the DC ports to the gold traces on the substrate. To reduce microwave losses at these interconnects, the chip and substrate are precisely aligned, and the RF electrodes on the chip are positioned in close proximity to the substrate, minimizing the wire bond lengths. A comparison of the magnitude responses of an on-chip MZM and PD before and after packaging is shown in Fig. S3(a) and S3(b), respectively. The degradation in the magnitude responses after packaging is primarily attributed to the wire bonding loss, the microstrip line loss, and the RF connector loss.

The post-packaging 3-dB bandwidths are 16 GHz for the MZM and 17 GHz for the PD.

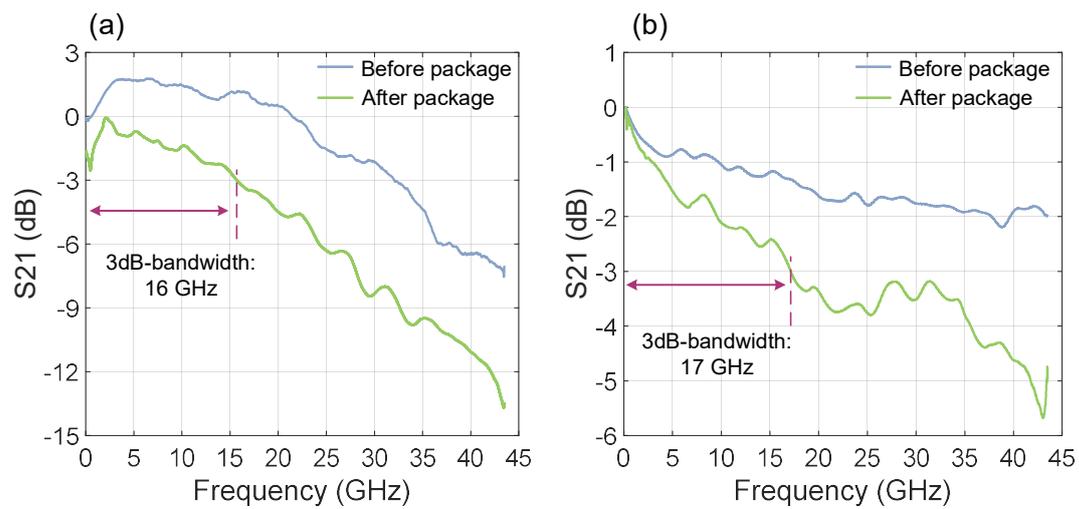

**Figure S3**. On-chip components characterization before and after packaging. (a) Measured magnitude response of an on-chip MZM. (b) Measured magnitude response of an on-chip PD. Blue curve: Pre-packaging measurement using RF probes. Green curve: Post-packaging measurement.

# S3. Calibration of the Optical Switch Delay Lines

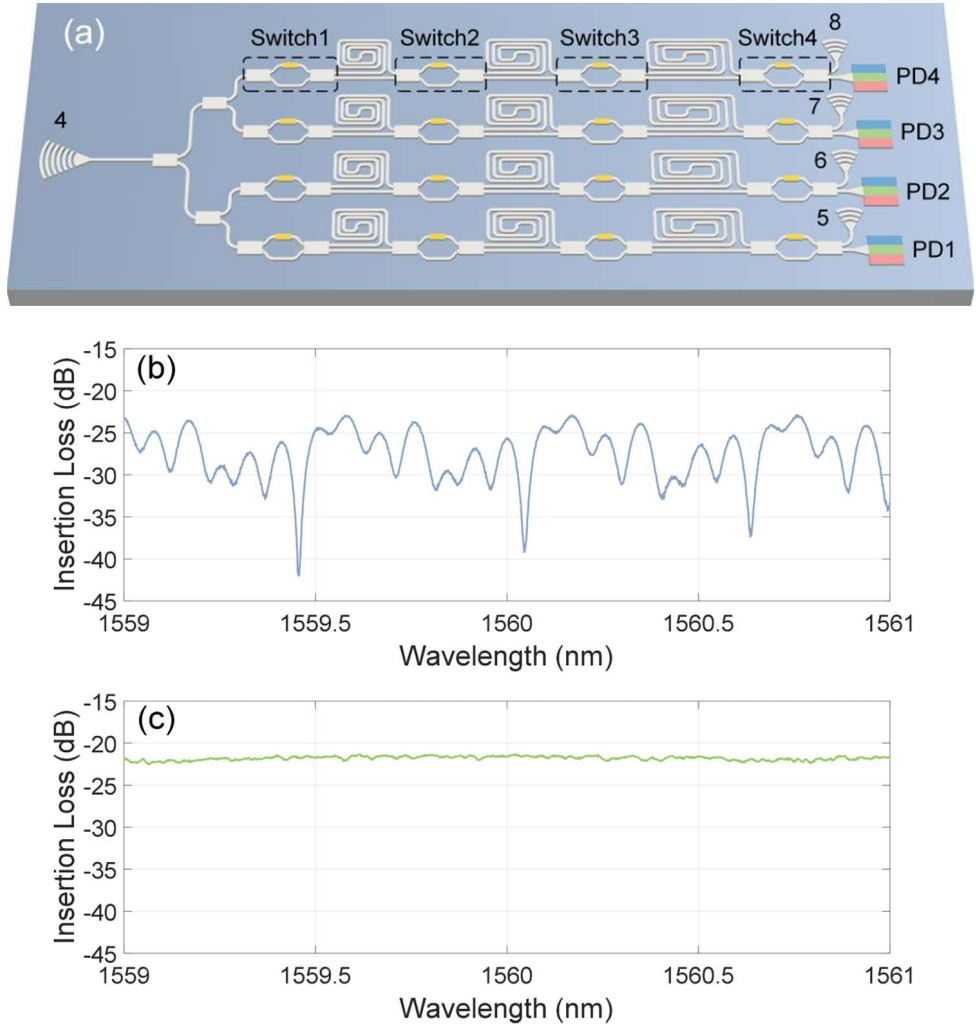

**Figure S4.** (a) The diagram of the microwave TTD beamforming network. (b) Optical spectral response measured before calibration. (c) Optical spectral response measured after calibration.

  In the design of the microwave TTD beamforming network, all switches are in the cross states when no voltages are applied. However, due to fabrication errors, the switches are not exactly in the cross states. When measuring the spectral response of the upper OSDL between the output port (port 8) and the input port (port 4), as shown in Fig. S4(a), the spectral response has variations due to optical interferences, as shown in Fig. S4(b). A solution to eliminate the errors is to perform calibration. Following the calibration method proposed in [1], we tune the voltages applied to switches 1~4 to allow all switches to operate in the cross states, which In the design of the microwave TTD beamforming network, all switches are in the cross states when no voltages are applied. However, due to fabrication errors, the switches are not exactly in the cross states. When measuring the spectral response of the upper OSDL between the output port (port 8) and the input port (port 4), as shown in Fig. S4(a), the spectral response has variations due to optical interferences, as shown in Fig. S4(b). A solution to eliminate the errors is to perform calibration. Following the calibration method proposed in [1], we tune the voltages applied to switches 1~4 to allow all switches to operate in the cross states, which results in a flat spectral response, as shown in Fig. S4(c).

The extinction ratios of the OSDLs in the microwave TTD beamforming network are also measured. To measure the extinction ratio of the upper OSDL, for example, all the switches in the OSDL are firstly calibrated to be in the bar state. The spectral response is shown as the solid black line in Fig. S5. Then, we change the state of one switch from bar state to cross state and measure the spectral response. For the four switches, four spectral responses are obtained, which are shown by the red, blue, green and pink lines in Fig. S5. As can be seen, the extinction ratio exceeds 20 dB over a 2 nm bandwidth.

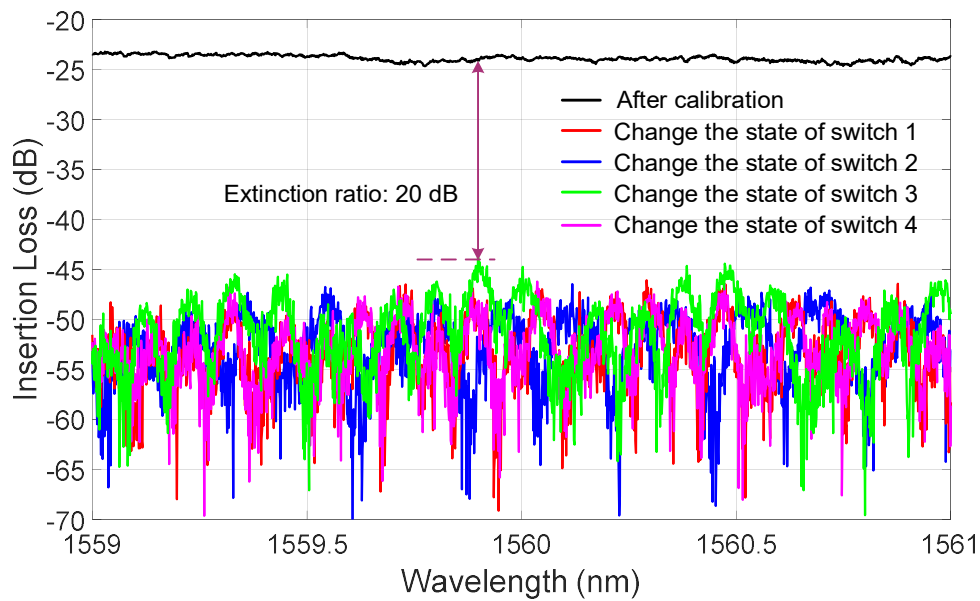

**Figure S5.** The extinction ratio measurements of the OSDL from part 4 to port 8.

# S4. Delay Measurement of the Microwave True time delay Beamforming Network

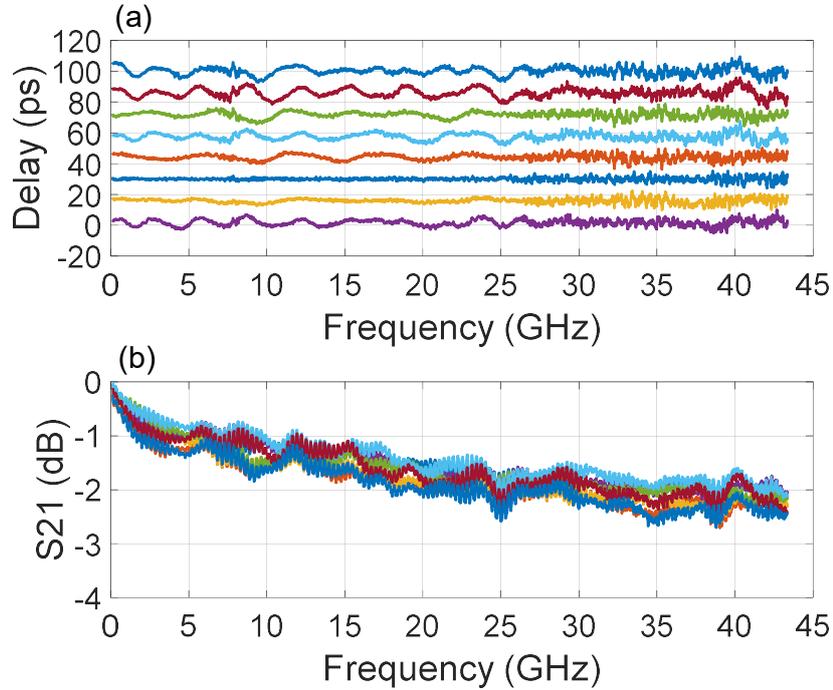

**Figure S6.** The measured (a) delay and (b) magnitude response of the OSDL.

To measure the time delays of the upper OSDL, an optical carrier modulated by a microwave signal is sent to the chip from port 4 via a grating coupler. The time delayed optical signal is applied to PD8. By changing the states of the optical switches, 8 different time delays are obtained. To maintain consistent optical input power to PD8 across these different time delays, the voltage applied to the fourth-stage optical switch is adjusted. This compensates for the insertion loss variation caused by the different waveguide lengths associated with each specific time delay. Figure S6(a) shows the measured time delays. As can be seen, the delay lines can provide 8 tunable time delays from 0–98 ps with a tuning step of 14 ps over a wide frequency range from 10 MHz to 43.5 GHz. The magnitude responses of the delay lines for different time delays are also measured, as shown in Fig. S6(b). The difference between the magnitude responses is below 1 dB.

# S5. The Calibration of the Optical Phased Array

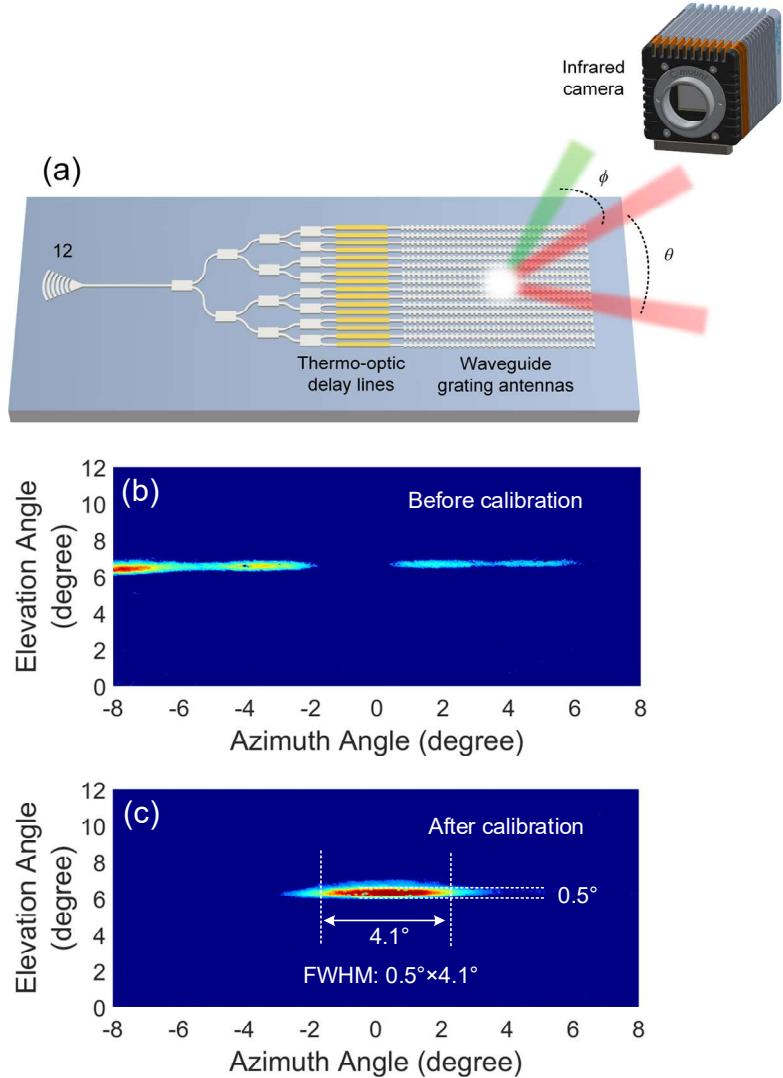

**Figure S7.** (a) The diagram of the OPA. (b) Optical spot measured before calibration. (c) Optical spot measured after calibration.

In the OPA system, fabrication-induced variations introduce initial delay mismatches among the sixteen waveguide delay lines, preventing the formation of a well-focused laser beam due to unmet phase-matching conditions. Therefore, calibration is needed. During the calibration, a continuous-wave (CW) light is coupled into the chip from port 12 via a grating coupler. The voltages applied to the sixteen thermo-optic delay lines are adjusted to minimize the initial delay difference, while an infrared (IR) camera is used to monitor the intensity profile of the beam in real time, as shown in Fig. S7(a). The calibration is deemed complete when the beam reaches maximum energy concentration. As seen in Fig. S7(b), without calibration, the beam energy is dispersed into multiple spots, indicating significant energy divergence. After calibration, as shown in Fig. S7(c), the energy converges into a spot with a full-width at half-maximum (FWHM) of 0.5° × 4.1°. By increasing the number of delay lines, the FWHM of the optical beam could be further reduced [2].

# S6. Characterization of the Optical Coherent Transmitter and Receiver

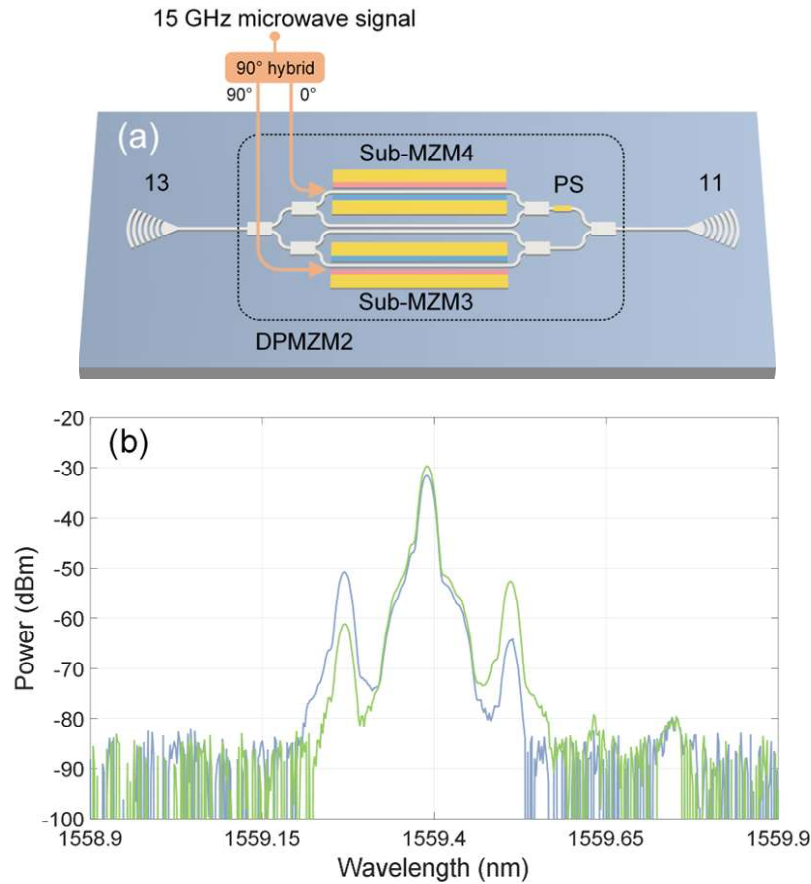

**Figure S8.** (a) The diagram of DPMZM2. (b) The measured optical spectra of the single sideband with carrier (C+SSB) modulation at port 11.

    The optical coherent transmitter employs a dual-parallel Mach-Zehnder modulator (DPMZM2), comprising two sub-Mach-Zehnder modulators (sub-MZM3 and sub-MZM4) and a thermal phase shifter (PS), as shown in Fig. S8(a). To verify the operation of the DPMZM2, a CW light is coupled into the chip from port 13 via a grating coupler. A 15 GHz microwave signal generated by an external microwave source is input to a 90° hybrid coupler, producing two microwave signals with a 90° phase difference. These two microwave signals are applied to sub-MZM3 and sub-MZM4 respectively. The voltage applied to the PS is adjusted to introduce a $\pi/2$ phase difference between upper and lower branch of DPMZM2 to realize single sideband with carrier (C+SSB) modulation. The optical spectra are measured at port 11, and are shown in Fig. S8(b). The blue line and the green line show the optical spectra of C+SSB modulation when a $\pi/2$ phase difference and a $-\pi/2$ phase difference are introduced, respectively. The results shown in Fig. S8(b) demonstrate that DPMZM2 can work properly.

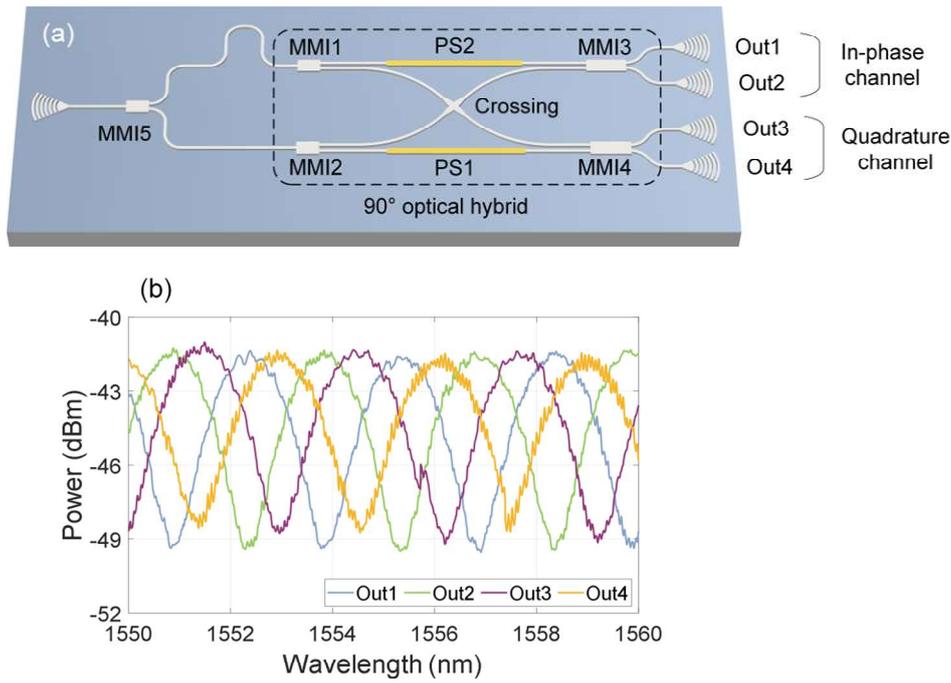

**Figure S9.** (a) The schematic diagram of a test 90° optical hybrid. (b) The measured spectral responses at the four output ports of the 90° optical hybrid.

The optical coherent receiver consists of two key components: a 90° optical hybrid and two balanced photodetectors (BPDs). To test the 90° optical hybrid, a separate 90° optical hybrid is designed, as shown Fig. in S9(a). The 90° optical hybrid is implemented using two 1×2 multimode interferometers (MMI1 and MMI2), two 2×2 MMIs (MMI3 and MMI4), a waveguide crossing and two thermal phase shifters (PS1 and PS2), while the BPDs are implemented using two discrete PDs with external circuit designed to subtract the output signals from the two PDs. To characterize the performance of the 90° optical hybrid, a broadband optical signal generated by an optical vector analyzer (OVA) is coupled into the chip via a grating coupler. This signal is split into two equal parts via a 3-dB optical coupler (MMI5) and fed into the hybrid, as shown in Fig. S9(a), with a controlled optical path delay difference between the two paths. Figure S9(b) presents the measured optical spectra at the four output ports of the hybrid, exhibiting sinusoidal interference patterns due to the phase difference induced by the delay. By tuning the two PSs within the hybrid, we enforce the required quadrature phase relationship between the outputs. The results clearly demonstrate the orthogonality between the in-phase (Out1 & Out2) and quadrature (Out3 & Out4) paths and also demonstrate a π-phase shift within the in-phase pair (Out1 vs. Out2) and the quadrature pair (Out3 vs. Out4). These measurements confirm the proper operation of the 90° optical hybrid, validating its capability for coherent detection.

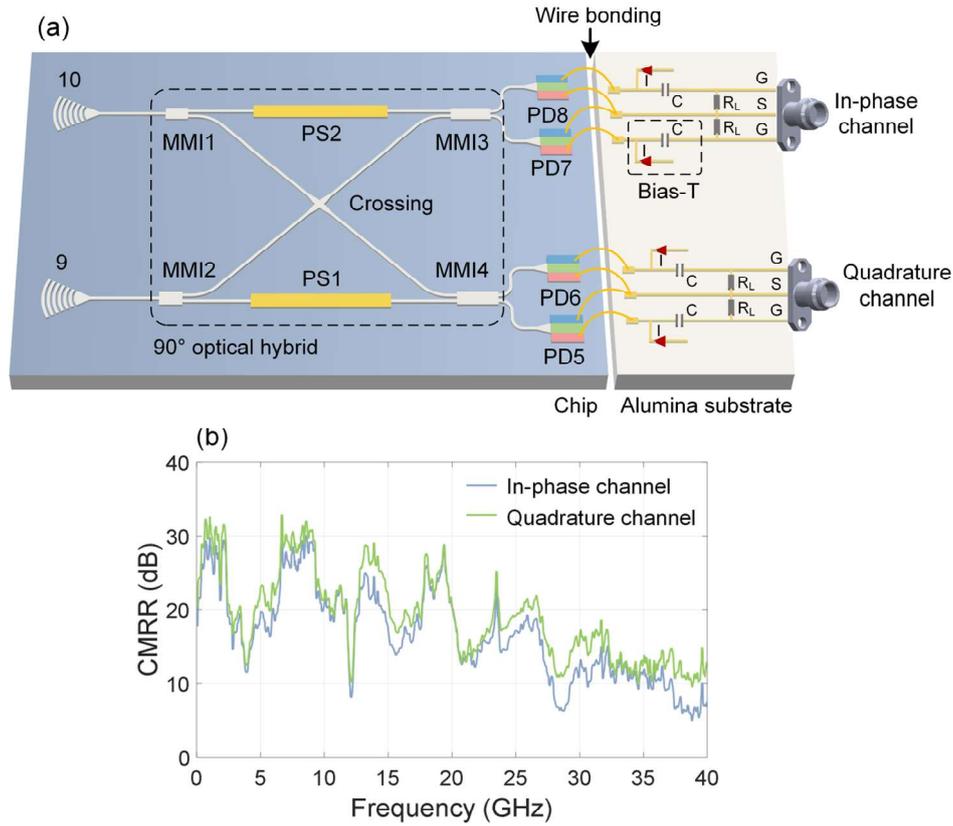

**Figure S10.** (a) The schematic diagram of the optical coherent receiver. (b) the measured common-mode rejection ratio (CMRR). I: inductance, C: capacitance, $R_L$: load resistance, G: ground, S: signal.

In the packaged modules, the BPDs are implemented on the chip and the outputs are connected to the microstrip lines on an external alumina substrate as shown in Fig. S10(a), which is designed according to [3]. To verify the functionality of these two BPDs, an optical carrier modulated with a microwave signal is coupled into the chip through port 10. After passing through the 90° optical hybrid, the optical signal is split into four parts and applied to PD5, PD6, PD7, and PD8. The signal of the in-phase channel is obtained by subtracting the signals from PD5 and PD6 using a circuit on the alumina substrate. Similarly, the signal of the quadrature channel is obtained by subtracting the signals from PD7 and PD8 using a circuit on the alumina substrate. The measured common-mode rejection ratio (CMRR) of the in-phase channel and quadrature channel are shown by the blue curve and green curve in Fig. S10(b), respectively. The CMRR exhibits significant degradation around 4 GHz and 12 GHz, which could be attributed to impedance mismatch in the external circuitry. Excluding the dip at 12 GHz, the bandwidth with CMRR >10 dB reaches 27.5 GHz. The experimental results shown in Fig. S10(b) demonstrate that these two BPDs can work properly.